\documentclass[sigconf,nonacm]{acmart}
\usepackage{float}
\usepackage{longtable}
\usepackage{booktabs}
\usepackage{listings}
\usepackage{amsmath}
\usepackage{makecell}
\usepackage{graphicx}
\usepackage{adjustbox}
\usepackage{algorithm}
\usepackage{algorithmic}
\usepackage{color}
\usepackage{placeins}
\usepackage{booktabs}
\usepackage[T1]{fontenc}
\AtBeginDocument{%
  }

\begin{document}

\title{Measuring Privacy vs. Fidelity in Synthetic Social Media Datasets}

\author{Henry Tari}
\orcid{0009-0009-5076-8732}
\affiliation{%
 \institution{Maastricht University}
 \city{Maastricht}
  \country{Netherlands}}
  \email{rouhollah.tari@maastrichtuniversity.nl}

\author{Adriana Iamnitchi}
\orcid{0000-0002-2397-8963}
\affiliation{%
  \institution{Maastricht University}
  \city{Maastricht}
  \country{Netherlands}}
  \email{a.iamnitchi@maastrichtuniversity.nl}



\begin{abstract}
  Synthetic data is increasingly used to support research without exposing sensitive user content. 
  Social media data is one of the types of datasets that would hugely benefit from representative synthetic equivalents that can be used to bootstrap research and allow reproducibility through data sharing. 
  However, recent studies show that (tabular) synthetic data is not inherently privacy-preserving. 
  Much less is known, however, about the privacy risks of synthetically generated unstructured texts. This work evaluates the privacy of synthetic Instagram posts generated by three state-of-the-art large language models 
  using two prompting strategies. 
  We propose a methodology that quantifies privacy by framing re-identification as an authorship attribution attack. A RoBERTa-large classifier trained on real posts achieved 81\% accuracy in authorship attribution on real data, but only 16.5--29.7\% on synthetic posts, showing reduced, though non-negligible, risk. Fidelity was assessed via text traits, sentiment, topic overlap, and embedding similarity, confirming the expected trade-off: higher fidelity coincides with greater privacy leakage. This work provides a framework for evaluating privacy in synthetic text and demonstrates the privacy--fidelity tension in social media datasets.
\end{abstract}


\maketitle
\section{Introduction}

The widespread use of social media platforms has produced vast amounts of user generated data that are highly valuable for research, industry, and policy making. Social media datasets enable insights into human behavior, public opinion, cultural trends, and more. However, they also contain sensitive personal information that can expose users to significant privacy risks if shared without proper safeguards. For this reason, most major platforms forbid researchers to publicly share social media datasets they collected for research, restricting sharing to data identifiers (such as tweet IDs) that theoretically enable the recollection of data. In reality, reliable and useful data recollection is often challenged by access to data collection API, removal of content by users or platforms, or algorithmic biased samples to data collection that, for example, favors new or popular content. 

In many fields, privacy constraints and limited access to real data have been addressed via synthetically generated datasets. By generating artificial but realistic data, researchers aim to preserve the statistical properties of the original dataset while reducing exposure of sensitive information. However, synthetic data are not inherently privacy-preserving. Recent studies~\cite{Stadler2020SyntheticD, Sarmin2025} have demonstrated that synthetic tabular data can remain vulnerable to re-identification, linkage, and inference attacks. These findings highlight the need for systematic and rigorous privacy evaluations of synthetic datasets rather than assuming safety based on data provenance.  

Most existing research on synthetic data privacy focuses on structured domains such as electronic health records~\cite{Chen2025-tf}, census data~\cite{Little_et.all}, and finance~\cite{potluru2024syntheticdataapplicationsfinance}, where privacy risks are well-established and quantifiable. By contrast, unstructured modalities such as text and, especially, short-form social media posts remain under explored. This gap is critical, as social media text is often distinctive in style, and linguistic patterns can function as implicit identifiers that undermine anonymity.
In particular, prior work primarily examined memorization and membership inference in large language models (LLMs)~\cite{yu2022,274574}, but has not systematically considered de-anonymization through authorship attribution for synthetic social media datasets. Authorship attribution models, originally developed in forensic linguistics and fraud detection, can exploit stylistic cues to link anonymous texts to individuals. In the context of synthetic social media datasets, this represents a direct and severe privacy risk: if synthetic posts can still be traced back to their authors, the intended anonymity is compromised.

At the same time, synthetic data must maintain fidelity to be useful for downstream research. This introduces a tension: increasing fidelity may preserve stylistic features that also aid re-identification, while increasing perturbations to improve privacy may reduce data quality. Despite its importance and the vast amount of work on the tension between utility and privacy in tabular~\cite{Stadler2020SyntheticD} and network data~\cite{Northam_2024}, this trade-off between privacy–fidelity in synthetic social media text remains largely unexplored.

Unlike other contexts, such as healthcare or finances, in which synthetic data is being used, social media data are typically public: the vast majority of research in computational social science relies on social media posts that are publicly available. The obvious question is thus: If the original data is public, why bother to protect its synthetic counterpart? There are multiple reasons for such concerns. First, posts that are easily attributed based on their synthetic representation can guide an adversary to an authorship identification attack. This can have dire personal consequences~\cite{Northam_2024} in autocracies or other contexts that penalise online speech.  Second, synthetic datasets may include representations of removed posts. Even if the original has been deleted, a synthetic surrogate can bring it ‘back’, undermining data subjects’ expectations of removal. The reidentification of authors of such removed posts based on the records in the synthetic dataset triggers violations of the ``right to be forgotten'' present in some legal systems around the world~\cite{GDPR}. 

This paper evaluates the privacy of synthetic social media data generated with large language models (LLMs). Specifically, this work introduces a methodology for quantifying the privacy exposure of synthetic textual data under de-anonymization attacks, using authorship attribution as the evaluation instrument. In addition, it evaluates the tension between the fidelity of synthetic social media datasets vs. privacy exposure. We evaluate fidelity through multiple dimensions, including social media traits such as emojis, tags, and post length, sentiment, topics, and embedding similarity.
Through these contributions, to our knowledge this paper provides the first systematic investigation of authorship attribution as a de-anonymization attack on synthetic social media text and offers insights into the trade-offs between fidelity and privacy in this area.

After reviewing related work on the privacy of synthetic data (Section~\ref{sec:related}), we describe the dataset used for our empirical evaluations (Section~\ref{sec:dataset}). Section~\ref{sec:synthetic} details our methodology for generating synthetic social media data using two prompts and a sample of the Instagram corpus. Section~\ref{sec:privacy} presents the authorship identification attack and its effectiveness on real data. In Section~\ref{sec:privacy-synthetic}, we discuss the privacy and fidelity of the synthetic data across LLMs and relate it to the observed privacy risks. Finally, Section~\ref{sec:conclusion} discusses our findings and their implications.

\section{Related Work}
\label{sec:related}

Synthetic data is seen as a promising solution to many challenges in data-driven research and practice. Several key motivations underlie its adoption across domains.
First, synthetic data helps address data scarcity in domains where access to real datasets is limited or subject to strict regulation~\cite{Mendes2025-vw}. 
Second, synthetic data serves as an effective form of data augmentation, particularly in contexts characterized by class imbalance. 
Third, synthetic datasets can be engineered to reduce bias and improve fairness, helping mitigate systematic biases in machine learning models~\cite{jordon2022syntheticdatawhat}.
Finally, synthetic data is a powerful tool for scalable training and simulation. 

Although synthetic data are often promoted as a privacy-preserving alternative to real datasets, recent studies show its lack of privacy guarantees~\cite{Stadler2020SyntheticD, Sarmin2025}: data generated to faithfully mimic real-world distributions may inadvertently encode sensitive details from the training corpus. 
Stadler et al.~\cite{Stadler2020SyntheticD} showed that synthetic data frequently repeats the same pitfalls as earlier anonymisation efforts. Their evaluation across multiple generative models revealed that non-private synthetic datasets remain vulnerable to linkage attacks, and even differentially private models face sharp trade-offs between privacy and utility. Importantly, they caution that privacy guarantees depend less on the label ``synthetic'' and more on rigorous design and evaluation.

Giomi et al.~\cite{giomi2022unifiedframeworkquantifyingprivacy} introduced \textit{Anonymeter}, a unified framework to assess privacy risk in synthetic tabular datasets. This tool quantifies risks such as singling-out, linkability, and inference, three core notions aligned with GDPR requirements. Their work highlights the need for principled risk assessment beyond ad-hoc or similarity-based metrics. Ganev and De Cristofaro~\cite{Ganev} demonstrated that popular similarity-based privacy metrics are inadequate, as synthetic datasets deemed “safe” by these metrics still leaked sensitive outlier records under their ReconSyn reconstruction attack.

Privacy risks in synthetic data are often assessed through adversarial attacks, including membership inference attacks~\cite{Shokri2017}, attribute inference attacks~\cite{Fredrikson2015}, linkage attacks~\cite{Narayanan} and reconstruction attacks~\cite{Ganev}.
A number of approaches have been proposed to mitigate these risks, each offering different trade-offs between utility, fidelity, and privacy. Differential privacy~\cite{DworkCalibrating,Abadi2016} enforces mathematical guarantees by introducing calibrated noise during model training. Differential privacy has been successfully applied to deep learning, including language models, where it can reduce the memorization of sensitive training data~\cite{yu2022}. However, it often degrades utility when applied aggressively, particularly for small or imbalanced datasets, making adoption challenging in practice. 

Adversarial training is another method for preserving privacy of synthetic data. Generative adversarial networks and variational autoencoders can be trained with privacy-aware objectives. For instance, discriminator networks can be designed to penalize the memorization of training samples, improving privacy at the cost of generation quality~\cite{xie2018differentially,Beaulieu-Jones2019-tf}. Such methods balance fidelity against leakage but lack the formal guarantees of differential privacy.

Post-processing and filtering is a pragmatic approach that removes synthetic records that are too similar to real individuals, often by applying nearest-neighbour distance thresholds or other similarity filters. While this reduces direct re-identification risk, it does not provide formal guarantees and can significantly affect fidelity. Tools such as \textit{Anonymeter}~\cite{giomi2022unifiedframeworkquantifyingprivacy} provide structured ways to evaluate the effectiveness of such post-processing.

Hybrid solutions combine classical anonymization with synthetic generation to enhance privacy protection in sensitive domains such as healthcare and finance~\cite{L-diversity,Stadler2020SyntheticD}. These methods can offer stronger protection but are often domain-specific and difficult to generalize.

Research on privacy in synthetic data has concentrated on structured and tabular domains such as electronic health records, census data, and finance~\cite{Stadler2020SyntheticD,giomi2022unifiedframeworkquantifyingprivacy}. 
Visual modalities such as facial images and medical imaging have also received growing attention. 
By contrast, textual synthetic data remains comparatively underexplored.

Several studies highlight both the promise and limitations of protecting privacy in textual synthetic data. 
Yu et al.~\cite{yu2022} introduced a framework for differentially private fine-tuning of large-scale language models. 
By leveraging parameter-efficient fine-tuning techniques, they showed that models such as RoBERTa and GPT-2 can maintain high accuracy under privacy constraints (e.g., only a 2--3\% drop compared to non-private baselines on MNLI and SST-2), setting a new standard in the privacy–utility trade-off. 
Building on this, Yue et al.~\cite{yue-etal-2023-synthetic} proposed an approach for synthetic text generation under differential privacy. 
They demonstrated that fine-tuning GPT-2 models with DP-SGD enables the generation of synthetic customer feedback and reviews with utility close to non-private models (within 2--4\% for classification tasks), while offering resilience against membership inference. 
Kurakin et al.~\cite{kurakin2024harnessinglargelanguagemodelsgenerate} refined this line by showing that parameter-efficient methods such as LoRA outperform full fine-tuning when training DP models for synthetic text, producing high quality data while mitigating privacy leakage from pre-trained models. 
Complementing these, Xie et al.~\cite{10.5555/3692070.3694313} introduced an API-based framework for generating differentially private synthetic data using foundation models, providing guarantees even when access to model internals is restricted. 

Other works emphasize risks. Carlini et al.~\cite{274574} demonstrated that large language models can memorize and regurgitate sensitive sequences from training corpora, underscoring the insufficiency of naive text generation for privacy. 
Similarly, Sarkar et al.~\cite{Sarkar2024-mp} compared de-identified and synthetic clinical notes, finding that membership inference attacks remain feasible even on synthetic datasets, thereby highlighting that privacy must be explicitly enforced rather than assumed. 

Textual data introduces unique challenges because linguistic style itself acts as an identifying signal. 
This makes \textit{de-anonymization through authorship attribution} a significant risk in synthetic social media datasets. 
While other attacks (e.g., attribute inference or membership inference) are possible, authorship attribution most directly undermines anonymity. 
The real-world implications are significant: in repressive regimes, individuals have faced severe punishment when anonymous writings were traced back to them~\cite{independentIranHangs}. 

To our knowledge, no prior work has systematically evaluated authorship attribution as a de-anonymization attack on synthetic social media text. This work addresses that gap, and borrows significantly from fidelity metrics reported in previous work~\cite{bertaglia2024Instasynth, Leveraging_GPT} on different datasets. 

\section{Dataset}
\label{sec:dataset}

Because our objective is to evaluate how easily is to identify the author of a social media posts given an observed collection of such posts, we used the Instagram collection from the Dutch Influencers Dataset introduced by Gui et al.~\cite{gui2024across}. 
The Instagram dataset consists of over \(116{,}000\) posts authored by 132 Dutch influencers. Posts from 2011 to 2023 were collected using the official Instagram API. Influencers often exhibit repetitive stylistic patterns tied to branding and niche topics, which are particularly valuable for authorship attribution tasks. The dataset is multilingual, with a nearly even split between Dutch (50.7\%) and English (49.3\%).

The multilingual composition of the dataset reflects realistic social media usage in non English contexts, where users frequently alternate between local language and English. To accommodate this setting, we use a multilingual transformer model for authorship attribution and sentiment analysis. This ensures that language variation does not artificially inflate or suppress attribution performance and allows us to evaluate privacy risks under realistic conditions.

\begin{figure}[htbp]
    \centering
    \includegraphics[width=1\linewidth]{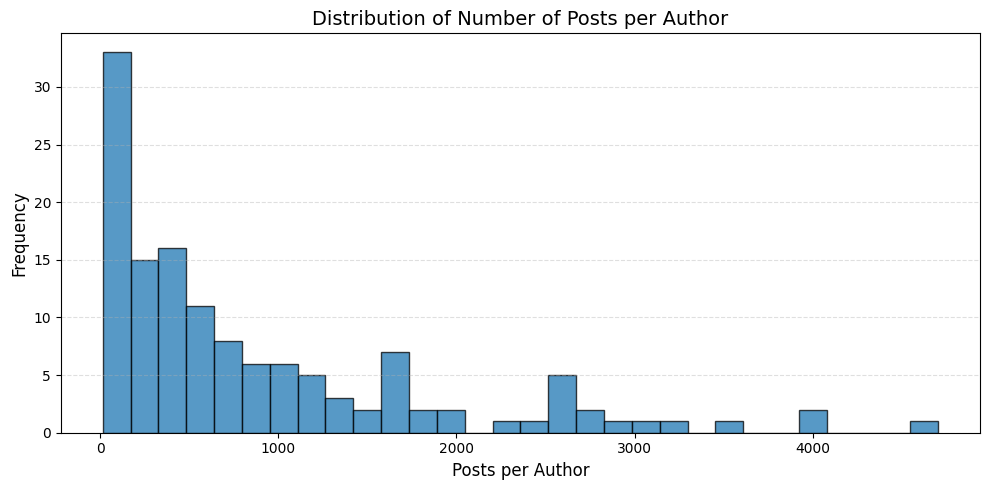}
    \caption{Distribution of number of posts per writer in the dataset. There are 30 bins with the size of 156.7. While the majority of authors have fewer than 1000 posts, some have over 4000, with a maximum of 6018.}
    \label{PostPerWriter}
\end{figure}

This dataset satisfies the central requirement for this study as it contain author information and multiple posts per author. This allows us to model authorship recognition by training a model to predict the author based solely on textual content. This dataset is particularly useful because it  consists of a large volume of posts generated by a relatively small number of users, which effectively simulates a worst-case privacy scenario.
The number of posts per user ranges from as few as 15 to as many as 6018. While most users contribute fewer than 1000 posts, a small number of prolific authors post several thousand times. Figure~\ref{PostPerWriter} shows the distribution of posts per user.

\begin{figure}[htbp]
    \centering
    \includegraphics[width=1\linewidth]{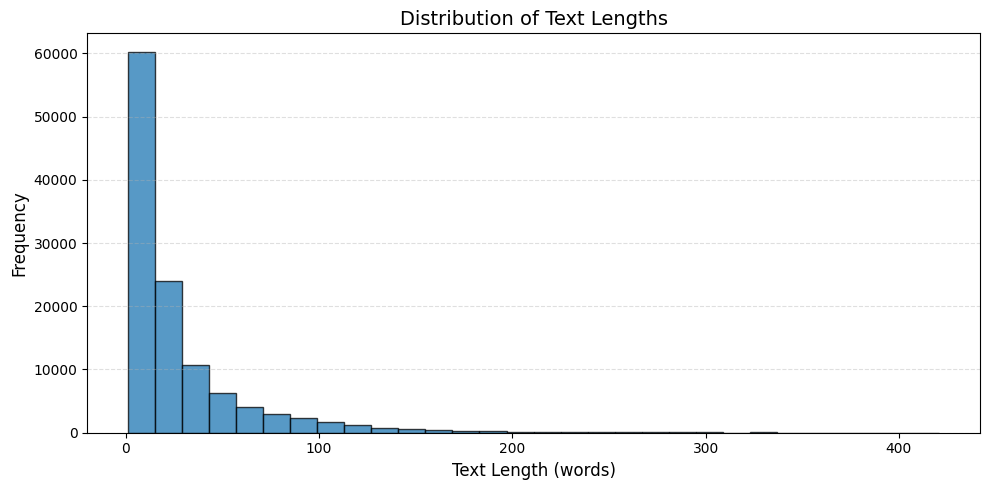}
    \caption{Distribution of post lengths. Most posts are shorter than 100 words with the median text length being 14. }
    \label{PostsLength}
\end{figure}

The average post length is relatively short, consistent with social media norms but shorter than desirable for textual analysis. Figure~\ref{PostsLength} shows that most posts contain fewer than 100 words.

The text content has been preprocessed by converting all characters to lowercase and removing formatting artifacts. Each record includes the author identifier, a timestamp, post text, hashtags, tagged users, and metadata such as like count, comment count, and flags for advertising or affiliate marketing.

For this study, we use the username and the post  text. Other metadata (such as subscriber count or like count) may provide valuable insights for future work but fall outside the scope of this paper. The post text includes free text as well as hashtags, user tags, emojis, and URLs, all of which may express aspects of the author’s personal style.

Despite its advantages, the dataset has some limitations. The average post length on Instagram is relatively short. However, this stylization does not reduce authorship signal; instead, influencer posts typically exhibit strong branding, recurring lexical choices, and consistent stylistic markers. Such features are particularly informative for stylometric models and therefore provide a stringent setting for evaluating authorship-based privacy risks. Additionally, the dataset is restricted to Dutch influencers, which may limit generalizability across broader social media populations. Nonetheless, its clear authorship structure and multilingual balance make it a strong fit for studying privacy in authorship attribution.

\section{Generating Synthetic Posts using LLMs}
\label{sec:synthetic}

To generate synthetic social media posts, we selected three state-of-the-art large language models (LLMs): OpenAI’s GPT-4o, Google’s Gemini 2.0 Flash and DeepSeek R1. These models were chosen based on their strong performance in standardized benchmarks, demonstrated few-shot generation capabilities, and robustness in handling informal and multilingual text characteristics critical for our synthetic data generation task.

\subsection{Identifying Representative Sample Size for Data Generation}

Generating synthetic versions of large datasets using different LLMs in multiple scenarios can quickly become prohibitively expensive. In our case, generating the entire dataset would have cost $~\approx \$1000$. This motivated the development of a strategy to generate synthetic data on a smaller, representative subset of the data. 

Several established formulas exist for determining representative sample sizes in numerical or tabular data. However, to the best of our knowledge, there are no established formulas that apply directly to unstructured text. To address this, we adapt statistical methodologies by transforming text into numeric vector representations and ensuring that the assumptions required by traditional formulas are reasonably satisfied. Specifically, we use the sample size estimation formula introduced by Cochran~\cite{cochran1977sampling}, which provides a mathematically grounded approach for estimating sample sizes in normally distributed data. The basic formula for sampling in infinite data space is:
\[
n = \frac{Z^2 \cdot \sigma^2}{E^2},
\]
where \( n \) is the required sample size, \( E \) is the error margin, \( Z \) is the Z-score corresponding to the desired confidence level, and \( \sigma \) is the standard deviation of the population. Taking into account the finite population size \( N \) of our dataset (approximately 116,000 posts), we apply Cochran’s adjusted formula:
\[
n = \frac{Z^2 \cdot \sigma^2 \cdot N}{E^2 (N - 1) + Z^2 \cdot \sigma^2}.
\]

To apply this method to text, we represent unstructured textual data using GloVe embeddings~\cite{pennington-etal-2014-glove}, which project each post into a 300-dimensional vector space. Embeddings provide a practical numerical representation of text, enabling the use of standard statistical methods. To compute the variance term \( \sigma^2 \), we use the trace of the covariance matrix, which aggregates variance across all dimensions into a single scalar measure:
\[
\sigma^2 = \text{Var}_{\text{total}} = \text{tr}(\Sigma).
\]

The assumption of normality is approximate rather than strictly satisfied. To check normality, we ran Shapiro–Wilk test on each dimension of our embedded dataset with W value in range of 0.9  and 0.99 suggesting normality.

We then distribute the total sample size across author strata using Neyman’s allocation~\cite{neyman1934}, which minimizes estimator variance by allocating samples proportionally to both stratum size and variability:
\[
n_w = n \cdot \frac{N_w \cdot \sigma_w}{\sum_{w=1}^{W} N_w \cdot \sigma_w},
\]
where \( N_w \) is the size of stratum \( w \), \( \sigma_w \) is the within-stratum standard deviation (computed via \( \text{tr}(\Sigma_w) \)), and \( W \) is the number of strata. Applying this method produced an initial sample size of \( n = 1{,}151 \). To ensure that all authors were represented, we rounded up each stratum’s allocation, yielding a final representative sample size of \( n = 1{,}216 \).

This adaptation of the Cochran method comes with limitations. First, we assume that embedding represents all aspect of textual data. While this can be said for context to some extent, it is not true for all aspect of texts. Second, by collapsing the covariance matrix into its trace, we treat total variance as a single scalar, discarding directional information about how variance is distributed across dimensions. Lastly, although Shapiro–Wilk results suggest near-normality, our reliance on large sample approximations introduces some uncertainty.
Taken together, these limitations mean that our approach should be regarded as a pragmatic approximation for representative sampling rather than a perfect statistical guarantee.

\subsection{Prompting Strategies}
To generate synthetic social media data using LLMs, we employed two distinct prompting strategies: \emph{Example-Based Prompting} and \emph{Persona-Based Prompting}.
Example-based prompting follows a conventional few-shot prompt design, where a series of real posts is provided as examples. The model is instructed to generate new posts that mimic the tone and structure of the examples. This approach builds directly upon techniques introduced in ~\cite{Leveraging_GPT}, and serves here as a high-fidelity baseline, closely aligned with the style and function of the original data.

Persona-based prompting introduces controlled stylistic variability while preserving semantic content. In this strategy, the LLM is instructed to assume the persona of a renowned 20th century literary figure and to rewrite social media posts in that author’s distinctive style. This strategy was inspired by the concept of k-anonymity. Just as k-anonymity seeks to make individual records indistinguishable within a group, persona-based prompting seeks to obscure author-specific stylistic markers by projecting them into alternative literary styles. In doing so, the generated posts retain their semantic meaning but become less attributable to the original author, potentially enhancing privacy. We deliberately selected 20th century literary figures because they represent stylistically distant and distinct writing profiles, thereby inducing maximal stylistic perturbation relative to contemporary social media language. The goal was not to approximate realistic influencer personas, but to evaluate whether strong stylistic displacement can reduce authorship signal. In this sense, persona-based prompting serves as an upper bound privacy heuristic: if substantial stylistic shifts fail to eliminate attribution, weaker stylistic adjustments are unlikely to do so.
Table~\ref{tab:prompt_templates} shows the prompt examples.

\begin{table}[t]
\centering
\caption{Prompt templates used in this study. As \emph{selected writers} in the persona-based propting strategies we used the authors specified in Table~\ref{tab:persona_intra_combined}.}
\label{tab:prompt_templates}
\begin{tabular}{p{\columnwidth}}
\toprule
\textbf{Example-Based Prompt Template} \\
\midrule
Using the following examples, generate 5 new social media posts, keeping them true to their content and writing style.

Please respond in the format shown below:

Post1: your response  
Post2: your response  
... \\
\midrule
\textbf{Persona-Based Prompt Template} \\
\midrule
You are \{selected writer\}, 
a renowned 20th-century literary figure known for your distinctive style.
Rewrite the following \{num examples\} social media posts in your own style,
preserving their meaning but adapting them to your voice.

Please respond in the format shown below:

Post1: your response  
Post2: your response  
... \\
\bottomrule
\end{tabular}
\end{table}

\section{The Privacy Risks of Social Media Data}
\label{sec:privacy}

In general, privacy risks in social media can be grouped into two broad categories: identity disclosure (re-identification) and attribute disclosure.  
Re-identification attacks exploit patterns, contextual cues, and auxiliary information to de-anonymize data and link it back to individuals.  
Attribute disclosure attacks aim to infer unknown personal attributes from available data, such as predicting a user’s home location~\cite{Home-Location} or demographic characteristics like age~\cite{Demographic-attributes}.  
The heterogeneous nature of social media data means that the applicable threats and defences vary depending on the modality: text, network structure, geolocation, or temporal signals. Comprehensive reviews of these risks are available in prior work~\cite{Privacy-in-Social-Media}.  

In this paper, we focus on re-identification attacks on textual data, as text serves both as the most pervasive feature of social media and as a powerful implicit identifier of individuals. This section evaluates the performance of different re-identification attacks on our original Instagram dataset. The best performing solution will then be tested on the synthetic data in Section~\ref{sec:privacy-synthetic}.

\subsection{Re-Identification Attack}

Re-identification attacks are a class of privacy breaches where the goal is to link anonymized data back to real individuals. Such attacks have been extensively studied in the context of health data, where adversaries exploit quasi-identifiers in supposedly anonymized records to re-identify patients~\cite{el2011systematic}. The methodology of a re-identification attack depends on both the data type and the anonymization strategy applied. 

For textual data, the most prominent re-identification approach is authorship attribution, which attempts to infer the author of an anonymous or synthetic text. While authorship attribution has a long history in forensic linguistics and fraud detection, its use as a privacy attack has received comparatively little attention. A related task, authorship verification, focuses on determining whether a text was written by a claimed author; this has been applied, for example, to detecting compromised accounts on social networks~\cite{Authorship-verification-Social-Network}. 

Authorship attribution methods can be broadly divided into two families: \emph{stylometric approaches} and \emph{language model approaches}, both of which have been studied extensively in forensic and security settings. Previous work has occasionally discussed their privacy implications, especially in the context of de-anonymization. In our work, we apply these techniques in a different way: we use them to assess privacy risks in synthetic social media datasets, using authorship leakage to estimate how much identifiability remains in the data.

\subsubsection{Stylometric Methods}

Stylometry is the study of measurable linguistic traits that characterize an individual’s writing style. Its history dates back centuries: in 1439, Lorenzo Valla used lexical comparisons to prove that the \emph{Donation of Constantine} was a forgery~\cite{20.500.11752/OPEN-141}. More recently, stylometry was famously applied by the FBI to link the \emph{Unabomber Manifesto} to Theodore Kaczynski~\cite{fbiUnabomberFederal}. 

Stylometric methods rely on extracting features that encode the writing style independently of content. These features can include basic lexical statistics (e.g. word length), structural measures (e.g. sentence length), and symbolic elements such as emojis or punctuation. In our experiments, we adopt a set of characteristics tailored to the characteristics of social media posts, combining established stylometric indicators with specific social media signals (e.g., hashtags, URLs, emojis). Table~\ref{tab:stylometric_features} summarizes the categories of features we extract.

\begin{table}[htbp]
\centering
\small
\setlength{\tabcolsep}{4pt}
\renewcommand{\arraystretch}{1.2}
\caption{Stylometric feature categories used in this study.}
\label{tab:stylometric_features}
\begin{tabular}{p{0.33\columnwidth} p{0.60\columnwidth}}
\toprule
\textbf{Feature Category} & \textbf{Description} \\
\midrule
Basic Text Statistics & Total words, sentences \\
Word Complexity & Average word length, lexical richness \\
Readability Scores & Flesch Reading Ease, Flesch–Kincaid Grade Level \\
Character-Level Counts & Digits, punctuation, special characters \\
Symbolic Elements & Emoji and emoticon count \\
Density Measures & Emoji/emoticon/hashtag/mention/URL densities \\
Repetition Features & Repeated words,emojis \\
Frequency Distributions & Frequency of letters, digits, and special characters \\
\bottomrule
\end{tabular}
\end{table}

\subsubsection{N-gram Methods}

N-grams capture recurring lexical and phrasal patterns that reflect an author’s stylistic habits. While unigrams capture word choice, bigrams and trigrams encode habitual collocations and local syntactic preferences that are often stable across topics.

This approach has a long tradition in authorship attribution. Keselj et al.~\cite{Keselj2003} introduced author profiles based on frequent n-grams, demonstrating robust performance across languages and domains. Sapkota et al.~\cite{Sapkota2015} later showed that character n-grams, particularly affixes and punctuation sequences, account for most of the discriminatory power in authorship-attribution models.

In our implementation, we extract the 100 most frequent bigrams and trigrams across the corpus. Each post is represented as a fixed-length feature vector indicating the frequency of these n-grams (with missing values zero-filled). We then train an $\ell_2$-regularized logistic regression classifier, using stratified splits by author. Although simple, n-gram baselines remain highly competitive for short, noisy text such as social media posts.

\subsubsection{TF-IDF Features}

Term frequency–inverse document frequency (TF-IDF) is a classical weighting scheme for identifying words that are frequent in one document but rare across the corpus. In authorship attribution, TF-IDF highlights distinctive vocabulary patterns while down-weighting common words. 

In our implementation, we extract unigram and bigram TF-IDF features with a vocabulary size capped at 3,000 tokens, balancing representation richness with computational efficiency. Posts are encoded as high-dimensional TF-IDF vectors, which are classified using $\ell_2$-regularized logistic regression. While TF-IDF cannot capture word order or long-range dependencies, it provides a transparent and competitive baseline for short-text attribution.

\subsubsection{Transformer-Based Methods}

Recent advances in transformer-based language models have significantly advanced authorship attribution. By fine-tuning pre-trained models such as BERT, researchers have shown that deep contextual embeddings capture subtle stylistic markers beyond surface-level lexical features. For example, the BertAA model~\cite{fabien-etal-2020-bertaa} fine-tunes BERT with a simple dense classification head, achieving up to 5.3\% relative improvement in accuracy on benchmark datasets such as Enron emails and IMDb62.

Building on this line of work, we fine-tune RoBERTa-large, a more recent model trained on larger corpora and optimized for better performance than vanilla BERT. We expect RoBERTa to be more effective on short, noisy texts typical of social media. We also incorporate social-media–specific stylometric features (e.g., emojis, hashtags, mentions) into our experimental comparisons, adapting the BertAA methodology to this domain. Finally, we perform hyperparameter tuning to identify configurations robust across varying numbers of authors.

Across our experiments, RoBERTa-large consistently outperformed stylometric, TF-IDF, and n-gram baselines. Ensembles of RoBERTa with engineered features provided only marginal gains, confirming the strength of transformer-based methods for this task. For consistency, all subsequent experiments use RoBERTa-large fine-tuned for 8 epochs with a batch size of 16 and a learning rate of $2 \times 10^{-5}$.

\subsection{Results of Re-Identification Attacks}

To assess how the difficulty of authorship attribution scales with the number of candidate authors, we first removed our sample from the dataset and ranked all remaining posts by Author's posting frequency. Based on this ranking, we constructed four evaluation subsets corresponding to the top 25\%, 50\%, 75\%, and 100\% most prolific authors. For each subset, we applied an 80/20 stratified split at the author level. This experimental design approximates realistic adversarial conditions in which an adversary can train authorship attribution models on large scale social media corpora. Importantly, we assume that the adversary does not have access to the original dataset from which the synthetic data are generated. Instead, it is assumed that the adversary trains the authorship attribution models on other publicly available data from the same platform. The released synthetic dataset is then treated as a new, previously unseen corpus whose authors the adversary attempts to identify using a model trained on external data. This reflects a realistic scenario in which a  well-resourced actor possesses large-scale platform data but not the specific dataset used to generate synthetic data. Because the platform environment is inherently multilingual, we assume that a well-resourced adversary would also train multilingual attribution models, as we do in our experiments. Under such conditions, an adversary could use a model trained on comprehensive datasets to re-identify users in a smaller, publicly released dataset containing sensitive or politically charged content.

\subsubsection{Single-Model Performance}

Table~\ref{tab:aa_results} summarize the results of the four individual models: TF-IDF, n-grams, stylometric features, and RoBERTa-large. Both accuracy and Macro-F1 are reported, since F1 is more informative under author imbalance.

\begin{table}[htbp]
\centering
\caption{Authorship attribution performance for individual models. Accuracy and Macro-F1 are reported.}
\begin{tabular}{|c|c|c|c|c|}
\hline
Subset (\%) & TF-IDF & N-grams & Stylometric & RoBERTa-large \\
\hline
25\%  & \makecell{Acc=71\% \\ F1=0.67} & \makecell{Acc=49\% \\ F1=0.44} & \makecell{Acc=43\% \\ F1=0.35} & \makecell{Acc=90\% \\ F1=0.88} \\
50\%  & \makecell{Acc=63\% \\ F1=0.52} & \makecell{Acc=41\% \\ F1=0.31} & \makecell{Acc=35\% \\ F1=0.23} & \makecell{Acc=85\% \\ F1=0.79} \\
75\%  & \makecell{Acc=59\% \\ F1=0.37} & \makecell{Acc=39\% \\ F1=0.23} & \makecell{Acc=32\% \\ F1=0.16} & \makecell{Acc=83\% \\ F1=0.69} \\
100\% & \makecell{Acc=58\% \\ F1=0.28} & \makecell{Acc=38\% \\ F1=0.18} & \makecell{Acc=32\% \\ F1=0.13} & \makecell{Acc=81\% \\ F1=0.58} \\
\hline
\end{tabular}
\label{tab:aa_results}
\end{table}

Three observations emerge. First, accuracy and F1 decline as the number of candidate authors increases, reflecting the greater difficulty in distinguishing among more writers. Second, among the feature-engineered baselines, TF-IDF performs the strongest, followed by n-grams, with stylometric features trailing. Third, RoBERTa-large achieves by far the best results, reaching 90\% accuracy (Macro-F1 = 0.88) on the top-25\% subset and maintaining 81\% accuracy (Macro-F1 = 0.58) even across the full author pool.

\subsubsection{Combined Models}

We also evaluated ensemble models that combine feature sets by stacking classifier probability outputs. This was motivated by the expectation that shallow lexical and stylistic features might provide complementary signals to deep contextual representations. Table~\ref{tab:aa_combined} presents the results.

\begin{table}[htbp]
\centering
\caption{Authorship attribution accuracy for combined models across author subsets.}
\begin{tabular}{|c|c|c|c|c|}
\hline
Subset & \makecell{TF-IDF \\ + Stylometric} & \makecell{RoBERTa \\ + TF-IDF} & \makecell{RoBERTa \\ + Stylometric} & \makecell{All \\ Combined} \\
\hline
25\%  & 0.76 & 0.77 & 0.61 & 0.81 \\
50\%  & 0.66 & 0.66 & 0.45 & 0.71 \\
75\%  & 0.61 & 0.60 & 0.41 & 0.64 \\
100\% & 0.60 & 0.56 & 0.39 & 0.60 \\
\hline
\end{tabular}
\label{tab:aa_combined}
\end{table}

The results show that combining TF-IDF and stylometric features yields the strongest ensemble among shallow models, slightly outperforming either feature alone. However, ensembles involving RoBERTa-large provide only marginal improvements and in some cases underperform relative to RoBERTa alone. The ``All Combined'' model reaches 81\% accuracy at 25\% authors but drops to 60\% accuracy at the full author set—no better than RoBERTa by itself. This indicates that RoBERTa already captures most of the discriminative lexical and stylistic signal, leaving little complementary information for shallow features to add.

Across all subsets and modelling strategies, RoBERTa-large consistently provides the best performance and scales more gracefully than the alternatives as the author pool grows. Although ensembles improve shallow models, they do not offer gains once a transformer model is included. Based on these findings, we adopt the RoBERTa-large configuration as the primary authorship attribution model for subsequent experiments.

\section{Privacy vs. Fidelity of the Synthetic Social Media Data}
\label{sec:privacy-synthetic}

The tension between privacy and utility in synthetic data has been extensively studied~\cite{Stadler2020SyntheticD,giomi2022unifiedframeworkquantifyingprivacy,Beaulieu-Jones2019-tf,8913684}. With the growing use of machine learning-based technologies, utility is increasingly seen as a task-specific measurement that sometimes is not in sync with the fidelity with which synthetic data replicates the real data. For example, Bertaglia et al.~\cite{bertaglia2024Instasynth} showed an instance where lower fidelity of an LLM-generated social media dataset led to higher utility for the detection of undisclosed ads. Because of this distinction between fidelity as an invariant that captures the relationship between synthetic and real data, and utility as a task-specific metric that measures the usefulness of the synthetic data, we adopt the fidelity metrics used in~\cite{Fidelity}.  
Thus, in this section we apply the best-performing authorship attribution model from the previous section, RoBERTa-large, to synthetic datasets generated by different LLMs and prompting strategies, and report the relationship between resistance to identification attacks and the fidelity of each synthetically generated dataset. 

\subsection{Re-Identification Results}

We used the RoBERTa-large configuration (8 epochs, batch size 16, learning rate $2\times 10^{-5}$) to perform authorship attribution attacks on all synthetic datasets. Table~\ref{tab:privacy_vs_original} summarizes privacy gains relative to the baseline of real data (81\% accuracy).

\begin{table}[htbp]
\centering
\caption{Authorship attribution accuracy on synthetic datasets and corresponding drops relative to the real-data baseline (81\%).}
\begin{tabular}{lcc}
\hline
\textbf{Model \& Prompt} & \textbf{Accuracy} & \textbf{Rel. reduction} \\
\hline
Gemini (example)   & 29.7\% & 63.33\% \\
Gemini (persona)   & 22.8\% & 71.85\% \\
DeepSeek (example) & 21.4\% & 73.58\% \\
DeepSeek (persona) & 16.5\% & 79.63\% \\
GPT-4o (example)   & 21.0\% & 74.07\% \\
GPT-4o (persona)   & 22.2\% & 72.59\% \\
\hline
\end{tabular}
\label{tab:privacy_vs_original}
\end{table}

Two clear trends emerge. First, authorship attribution becomes substantially harder on synthetic data: average accuracy drops from 81\% on real posts to roughly 25\% on synthetic data. This corresponds to relative reductions of 63--80\%, depending on model and prompting strategy. Second, persona-based prompting tends to yield stronger privacy protection than example-based prompting, with DeepSeek showing the largest improvement. The exception is GPT-4o, where persona-based prompting performs slightly worse.

Importantly, although synthetic generation significantly reduces re-identification risk, the attack does not vanish entirely: accuracies around 20--30\% remain well above random chance. This indicates that synthetic text still retains detectable stylistic traces of authorship, underscoring the need to evaluate both privacy and fidelity in tandem.

\subsection{Fidelity Evaluation}

For fidelity evaluation, we follow Tari et al.~\cite{Fidelity} in measuring how each LLM compares under different prompting strategies. We assess fidelity along several dimensions: (i) social media traits and textual characteristics, (ii) sentiment distribution, (iii) topical coverage and overlap, and (iv) embedding-based similarity to real data.

\subsubsection{Text Characteristics and Social Media Traits}

A central dimension of fidelity is the extent to which the generated text reproduces characteristic features of social media writing. These include hashtags, mentions, URLs, and emojis, as well as broader textual properties such as text length, lexical diversity, readability, sentence length, and punctuation usage. Readability was measured using the Flesch Reading Ease score~\cite{flesch-1948}, and lexical diversity was calculated following~\cite{LExical_Diversity}. Table~\ref{tab:model_stats} reports average values across models and prompting strategies, compared to the original dataset.

\begin{table*}[htbp]
\centering
\caption{Average social media trait statistics and textual characteristics per post for each model across prompting strategies, compared to the original dataset. Bold text marks the text characteristics closest to the original.}
\small
\begin{adjustbox}{max width=\textwidth}
\begin{tabular}{lrrrrrrrrr}
\toprule
Model             & Hashtags & Mentions & URLs & Emojis & Text Length & Readability & Lexical Diversity & Avg. Sentence Length & Punctuation\\
\midrule
DeepSeek          & 1.38 & \textbf{0.42} & \textbf{0.015}  & 1.19 & 17.21 & 62.31 & 0.98 & 7.57 & 4.68 \\
DeepSeek\_Persona  & 0.71 & 0.31 & 0.0033 & 0.16 & 25.21 & 61.37 & 0.95 & 7.04 & \textbf{7.48} \\
GPT-4o            & 1.41 & 0.28 & 0.0091 & \textbf{1.08} & 22.35 & 65.94 & 0.96 & 8.70 & 5.11 \\
GPT-4o\_Persona    & 0.85 & 0.33 & 0.0116 & 0.42 & 29.25 & 58.26 & \textbf{0.93} & \textbf{11.27} & 6.63 \\
Gemini            & \textbf{2.00} & 0.32 & 0.0115 & 0.67 & 19.50 & 67.93 & 0.97 & 6.55 & 6.03 \\
Gemini\_Persona    & 1.28 & 0.20 & 0.0058 & 0.09 & \textbf{34.85} & \textbf{64.57} & \textbf{0.91} & 7.00 & 12.49 \\
Original          & 2.89 & 0.74 & 0.0230 & 0.94 & 35.43 & 64.42 & 0.92 & 11.61 & 9.29 \\
\bottomrule
\end{tabular}
\end{adjustbox}
\label{tab:model_stats}
\end{table*}

Several patterns emerge: Counts of hashtags, mentions, and URLs are substantially lower in all synthetic datasets compared to the original corpus. The reduction is especially pronounced in persona-based prompting, which de-emphasizes such markers. Among models, Gemini best preserves these elements.
Emoji usage decreases across synthetic data, again more sharply under persona-based prompting. This suggests that stylistic transformation reduces alignment with social media conventions.
Generated posts are consistently shorter than originals in terms of word and sentence length. Persona-based prompting partially mitigates this gap, producing longer posts with richer punctuation. Scores remain broadly comparable to real posts across all conditions, with only minor decreases in persona-based outputs. This indicates that surface-level complexity and vocabulary variety are preserved, even as social-media–specific markers decline.

\subsubsection{Sentiment Analysis}

Another dimension of fidelity is how well synthetic posts preserve the sentiment distribution observed in the original data. For this analysis, we employed a state-of-the-art multilingual model fine-tuned on social media text~\cite{barbieri-etal-2022-xlm} to classify posts from both real and synthetic datasets.

\begin{figure}[htbp]
    \centering
    \includegraphics[width=1\linewidth]{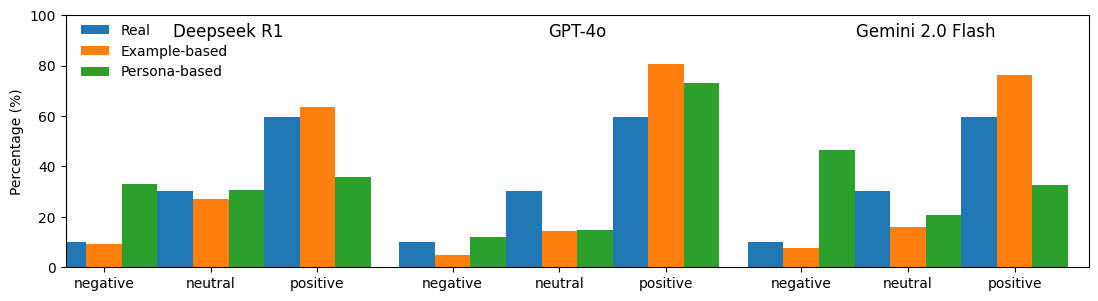}
    \caption{Sentiment distribution for all LLMs compared to real data under example-based and persona-based prompting. }
    \label{fig:sentiment}
\end{figure}

Figure~\ref{fig:sentiment} reports the sentiment distribution for all models. Under example-based prompting, GPT-4o replicates the tendency previously observed in GPT-3.5-turbo~\cite{Leveraging_GPT, Fidelity}: negative and neutral sentiments are reduced, while positive sentiment is amplified. In contrast, persona-based prompting shifts the distribution toward more negative sentiment and less positive sentiment, with neutral sentiment remaining largely stable.

For DeepSeek-R1, the example-based prompt yields sentiment distributions that closely mirror the real data, with only minor reductions in negative and neutral categories and a slight increase in positive sentiment. However, Persona-based prompting results in a substantial increase in negative sentiment and a corresponding decrease in positive sentiment.

Gemini 2.0 Flash follows a pattern similar to that of GPT-4o. Under example-based prompting, positive sentiment is exaggerated at the expense of negative and neutral categories. However, Persona-based prompting produces a more pronounced increase in negative sentiment and a decrease in positive sentiment.
Across models, a consistent pattern emerges. Example-based prompting generally exaggerates positive sentiment, though DeepSeek remains the closest to the real distribution. Persona-based prompting has the opposite effect, amplifying negative sentiment at the expense of positive sentiment.

While the previous analysis compares aggregate sentiment distributions, it does not capture whether individual posts retain their original affective label. To assess instance-level fidelity, we computed pairwise sentiment preservation between each original post and its synthetic counterpart. Specifically, for each model and prompting strategy, we measured the percentage of synthetic posts whose predicted sentiment matched that of the original post.

Table~\ref{tab:sentiment_preservation} reports the preservation rates per sentiment category. Results reveal asymmetric behavior across models and prompting strategies. Example-based prompting tends to strongly preserve positive sentiment (up to 90.21\% for GPT-4o and 89.10\% for Gemini), while preservation of negative and neutral sentiment is substantially lower. Persona-based prompting, by contrast, increases preservation of negative sentiment in some cases (e.g., 79.67\% for Gemini), but often reduces preservation of positive sentiment. These findings indicate that distributional similarity does not necessarily imply instance level fidelity, and that stylistic perturbation can selectively shift affective tone even when overall sentiment proportions appear comparable.

\begin{table}[htbp]
\centering
\caption{Pairwise sentiment preservation between original and synthetic posts. Values report the percentage of synthetic posts that retain the original sentiment label.}
\label{tab:sentiment_preservation}

\resizebox{\columnwidth}{!}{%
\begin{tabular}{llccc}
\toprule
\textbf{Model} & \textbf{Prompt} & \textbf{Negative (\%)} & \textbf{Neutral (\%)} & \textbf{Positive (\%)} \\
\midrule
DeepSeek & Example-based & 35.77 & 46.20 & 76.97 \\
DeepSeek & Persona-based & 64.23 & 45.11 & 47.31 \\
\midrule
GPT-4o   & Example-based & 17.89 & 26.36 & 90.21 \\
GPT-4o   & Persona-based & 54.47 & 28.26 & 86.76 \\
\midrule
Gemini   & Example-based & 33.33 & 32.34 & 89.10 \\
Gemini   & Persona-based & 79.67 & 32.88 & 41.38 \\
\bottomrule
\end{tabular}%
}
\end{table}

\subsubsection{Topic Modeling}

Topic modeling provides another lens for assessing the fidelity of synthetic data by examining whether generated posts reproduce the thematic structure of the original dataset. We employed BERTopic~\cite{grootendorst2022bertopic} with the \textit{paraphrase-multilingual-MiniLM-L12-v2} sentence transformer~\cite{reimers2020multilingual}. Model parameters were tuned using grid search, with topic coherence ($c_v$)~\cite{roder2015exploring} as the evaluation metric. The configuration with \texttt{min\_topic\_size}=10 produced the highest average coherence score of 0.4345 and was therefore selected for subsequent experiments, with \texttt{nr\_topics="auto"} and per-group refitting.

Using this configuration, we extracted topic vectors for each dataset and constructed cosine similarity matrices. The topics were matched between the real and synthetic sets using a greedy algorithm with a similarity threshold of 0.7. Shared topics represent thematic overlap, while unique topics indicate divergences introduced by the LLMs.

\begin{figure}[htbp]
    \centering
    \includegraphics[width=1\linewidth]{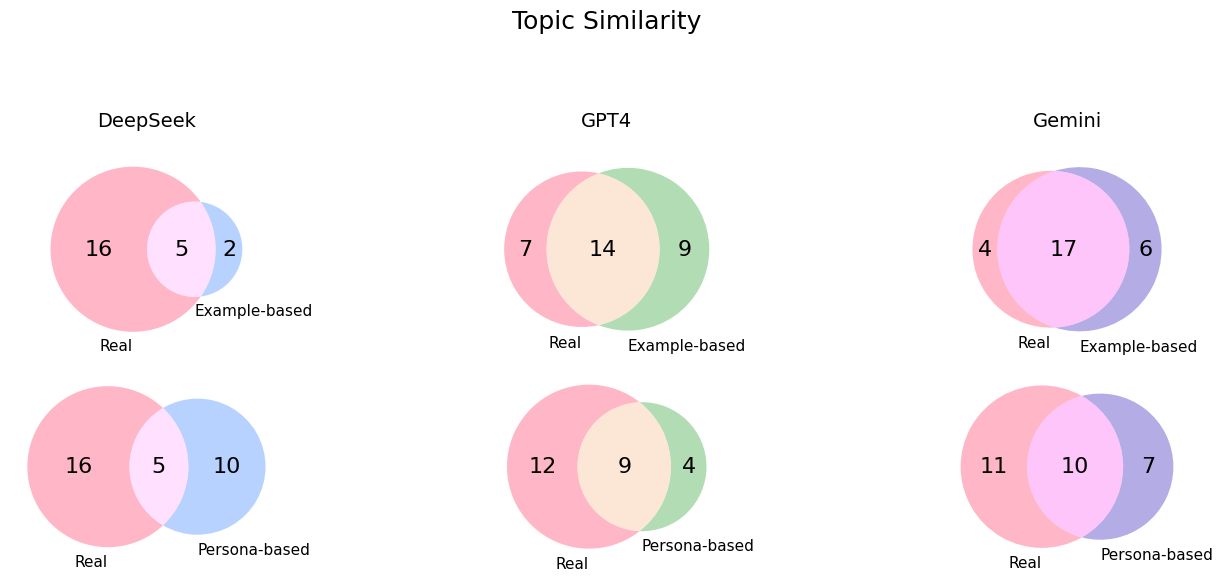}
    \caption{Topic overlap between real and DeepSeek-generated data. Shared topics remain stable across prompts, but persona-based prompting produces many more unique topics.}
    \label{fig:TS}
\end{figure}

Figure~\ref{fig:TS} shows the topic similarity results. In DeepSeek under example-based prompting, five topics were shared with real data and two unique topics emerged. Under persona-based prompting, the number of shared topics remained constant, but unique topics increased markedly to 10, indicating that DeepSeek generates more spurious topics when guided by literary personas. GPT's results shows that with example-based prompting, GPT-4o generated 14 shared and nine unique topics. Persona-based prompting reduced both: nine shared topics and four unique topics. This suggests that stylistic transformations dampen GPT-4o’s ability to reproduce topical variety. For Gemini, example-based prompting yielded 17 shared and six unique topics, while persona-based prompting produced 10 shared topics and a similar number of unique topics. Compared to other models, Gemini preserved more of the topical structure of real data.
Overall, Gemini demonstrated the strongest topical fidelity, preserving the largest number of shared topics. GPT-4o preserved fewer topics under persona-based prompting, while DeepSeek introduced many unique, low-overlap topics that likely reduce fidelity. 

\subsubsection{Embedding Space}
\begin{figure}[htbp]
    \centering
    \includegraphics[width=0.8\linewidth]{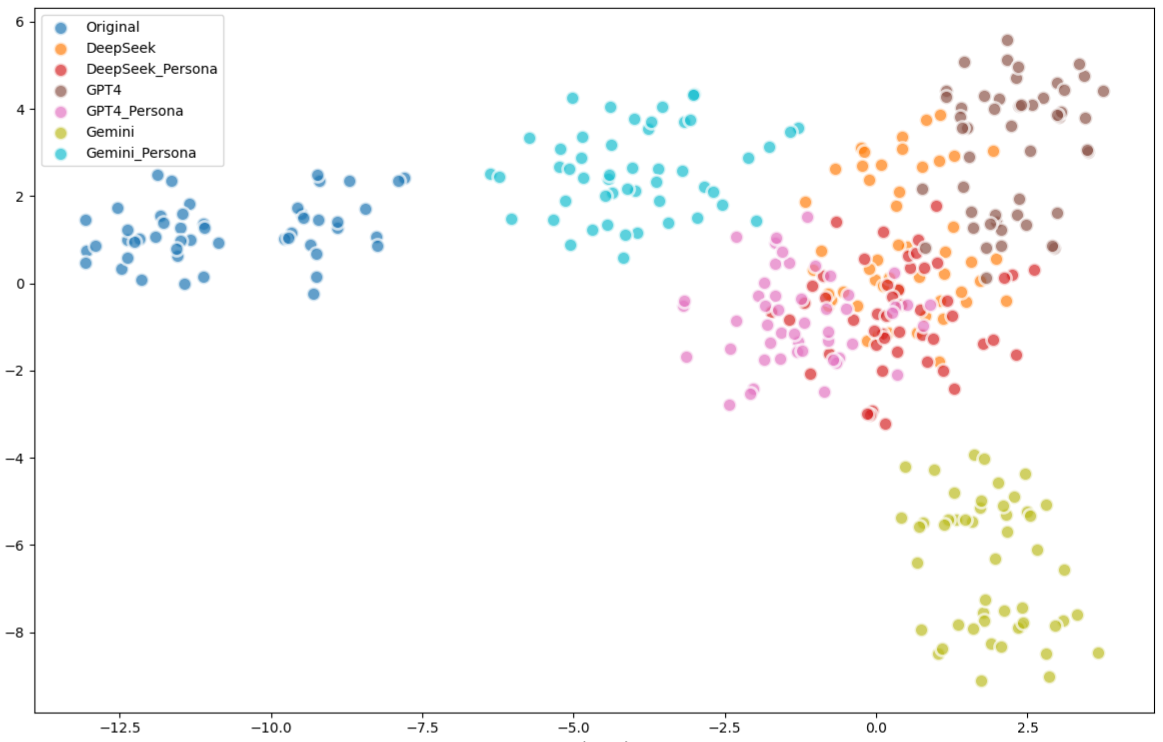}
    \caption{t-SNE visualization of generated posts by models per prompting strategy and original posts.}
    \label{fig:Embedding}
\end{figure}

While the embedding space is not directly a fidelity metric, it provides useful insights into why privacy outcomes differ between real and synthetic data. Figure~\ref{fig:Embedding} shows a t-SNE visualization of real and synthetic posts across LLMs and prompting strategies. For visualization, we first clustered all posts into 50 centroids using $k$-means and then projected these centroids into two dimensions with t-SNE. Although t-SNE distorts absolute distances, the fact that synthetic posts occupy distinct and distant regions from real posts helps explain why authorship attribution attacks behave so differently between the two.

As noted earlier, our inspiration for persona-based prompting comes from $k$-anonymity. The intuition is that if posts generated under the same literary persona are closer to each other, they become less distinguishable by the original author.  To see if it works, we  measured the average pairwise Euclidean distance of centroids for each writer and compared it to the global average centroid distance. Table~\ref{tab:persona_intra_combined} summarizes the results. 

\begin{table*}[htbp]
\centering
\setlength{\tabcolsep}{8pt}
\renewcommand{\arraystretch}{1.2}
\caption{Intra-writer clustering relative to model-level global centroid distances (Euclidean). Each cell reports $\Delta = d_{\text{intra}}(\text{writer}) - d_{\text{global}}(\text{model})$, where $d_{\text{intra}}$ is the mean pairwise distance between centroids for that writer's persona and $d_{\text{global}}$ is the model's mean pairwise distance over all centroids. Negative (bold) values indicate stylistically tighter writer clusters than the model's global baseline. Global means: DeepSeek = 0.4595, GPT-4 = 0.4340, Gemini = 0.4961.}
\label{tab:persona_intra_combined}
\begin{tabular}{lccc}
\toprule
\textbf{Writer} & $\boldsymbol{\Delta}$ \textbf{ DeepSeek\_Persona} & $\boldsymbol{\Delta}$ \textbf{ GPT4\_Persona} & $\boldsymbol{\Delta}$ \textbf{ Gemini\_Persona} \\
\midrule
Ernest Hemingway    & \textbf{-0.0232} & +0.0520          & \textbf{-0.0559} \\
F. Scott Fitzgerald & \textbf{-0.0511} & +0.0038          & +0.0329          \\
George Orwell       & \textbf{-0.0611} & +0.0365          & \textbf{-0.0292} \\
James Joyce         & \textbf{-0.0475} & +0.0992          & \textbf{-0.0088} \\
John Steinbeck      & +0.0146          & +0.0160          & \textbf{-0.0722} \\
Kurt Vonnegut       & \textbf{-0.0613} & \textbf{-0.0173} & +0.0012          \\
Samuel Beckett      & \textbf{-0.0064} & \textbf{-0.0257} & \textbf{-0.0922} \\
T.S. Eliot          & \textbf{-0.0772} & \textbf{-0.0133} & \textbf{-0.0218} \\
Virginia Woolf      & \textbf{-0.0537} & \textbf{-0.0287} & \textbf{-0.0536} \\
William Faulkner    & \textbf{-0.0716} & \textbf{-0.0481} & \textbf{-0.0386} \\
\bottomrule
\end{tabular}
\end{table*}

A negative $\Delta$ indicates that intra-writer centroids are closer together than the global mean distance. We see that DeepSeek and Gemini made posts closer together for nine and eight writers out of 10, respectively, while GPT-4o behaved almost randomly (five closer vs.\ five farther). This may explain why GPT-4o was the only model where persona-based prompting did not improve privacy.
Taken together, these findings indicate that fidelity degradation is uneven across dimensions. Example-based prompting preserves many structural features and topical distributions, but retains substantial stylistic proximity to the original authors. Persona-based prompting induces stronger stylistic displacement, as reflected in embedding centroid shifts and reduced attribution accuracy, but alters platform specific markers such as hashtag usage and stylistic conventions. Importantly, no configuration simultaneously achieves maximal privacy and complete fidelity preservation. The trade-off manifests differently across lexical, structural, and semantic dimensions, underscoring that fidelity is multi-faceted rather than monolithic.
\section{Conclusion}
\label{sec:conclusion}

This paper answers the questions: What are the author identification risks in LLM-generated synthetic social media datasets and how are these risks related to how accurate the synthetic data reflects the real data characteristics? We present a methodology for quantifying privacy risks in synthetic social media text by operationalizing re-identification as an authorship attribution attack. In contrast to prior work that predominantly employed similarity measures or utility metrics, the proposed approach trains an authorship attribution model on real-world data and subsequently evaluates its performance on synthetic posts. This enables an empirical, attack-driven assessment of anonymity.

Evaluating three large language models (GPT-4o, Gemini 2.0 Flash, DeepSeek R1) with example-based and persona-based prompting yielded several findings. The accuracy of authorship attribution fell from 81\% in real posts to 16.5\% in synthetic posts, indicating substantial but incomplete privacy protection. Our persona-based prompt inspired by k-anonymity improved privacy for DeepSeek and Gemini, but not for GPT-4o, where the clustering of embedding space was weak, suggesting that stylistic obfuscation depends on both the model architecture and the generation strategy.

At the same time, fidelity preservation was not uniform across prompting strategies or textual dimensions. While semantic content and topical distributions remained largely stable across models, stylistic and platform-specific markers shifted more substantially under persona-based prompting. These findings indicate that fidelity cannot be treated as a single scalar property, but rather as a multidimensional construct in which lexical, structural, and semantic characteristics respond differently to privacy-oriented transformations. The privacy–fidelity trade-off therefore manifests unevenly: stronger stylistic displacement may reduce attribution accuracy, but it can also alter surface-level features that contribute to platform authenticity.
This study has several limitations. First, the dataset contains only short-form Instagram posts, limiting generalisation to longer or structurally different content on platforms such as Reddit or Facebook. However, the methodological framework is, in principle, transferable to other platforms and formats.
Second, while persona-based prompting was explored as a privacy heuristic, established privacy-enhancing techniques (e.g., differential privacy, k-anonymity, t-closeness) were not empirically tested. Their comparative effectiveness relative to persona-based prompting therefore remains unknown.
Third, the analysis only considered textual content. In practice, social media data also include rich metadata such as timestamps, interaction, reply graphs, and multimodal content such as images and videos. These modalities introduce distinct privacy risks that were not evaluated here.
And finally, the adversarial evaluation focused solely on de-anonymization via authorship attribution. Other attacks such as membership inference, attribute inference, or data reconstruction were not examined and may expose additional or qualitatively different privacy vulnerabilities.

Building on this work, future research directions include extending the proposed methodology to multimodal platforms, thereby enabling the derivation of more generalizable insights. Another promising avenue involves the systematic evaluation of classical privacy-preserving mechanisms in conjunction with LLM prompting strategies, as well as the exploration of alternative text generation paradigms to achieve a more effective balance between fidelity and privacy. Future work should also encompass comprehensive investigations of additional adversarial attacks beyond authorship attribution, in order to obtain a more complete characterization of privacy risks. 

\bibliographystyle{plainnat}
\bibliography{ref}

\end{document}